\documentstyle[prd,aps,floats]{revtex}
\tighten

\begin{document}
\draft
\preprint{\ 
\begin{tabular}{rr}
 &CfPA 96-th-29  \\ 
& 
\end{tabular}
} 
\twocolumn[\hsize\textwidth\columnwidth\hsize\csname@twocolumnfalse\endcsname 
\title{Constraints on a Primordial Magnetic Field}
\author{John D. Barrow$^{1,2}$, Pedro G. Ferreira$^{1}$ and Joseph Silk$^{1}$ }
\address{$^{1}$Center for Particle Astrophysics,  and  Departments of Astronomy and Physics,\\
University of California, Berkeley  CA 94720-7304\\
$^{2}$Astronomy Centre, University of Sussex, Brighton BN1 9QH,
U.K.}
\maketitle
\begin{abstract}
We derive an upper limit of $B_0<3.4\times 10^{-9}(\Omega_0h_{50}^2)
^{\frac 12}$
Gauss on the present strength of any primordial homogeneous magnetic
field.
 The microwave background
anisotropy created by cosmological magnetic fields is calculated  in the most general flat
and open anisotropic cosmologies containing expansion-rate and 3-curvature
anisotropies.  Our limit is derived from a statistical analysis of the 4-year Cosmic
Background Explorer data for anisotropy patterns characteristic of
homogeneous anisotropy averaged over all possible sky orientations with
respect to the COBE receiver. The limits we obtain are considerably stronger
than those imposed by primordial nucleosynthesis and ensure that other
magnetic field effects on the microwave background structure are
unobservably small.
\end{abstract}
\date{\today}
] \renewcommand{\thefootnote}{\arabic{footnote}} \setcounter{footnote}{0}

The origin of magnetic fields observed in galaxies and galaxy clusters is
still a mystery. The invocation of protogalactic dynamos to explain the
magnitude of the field involves many uncertain assumptions but still
requires a small primordial (pregalactic) seed field \cite{a}. Hence the
possibility of a primordial field merits serious consideration. Other
attempts to find an origin for the field in the early universe have appealed
to battery effects, the electroweak phase transition, or to fundamental
changes in the nature of the electromagnetic interaction. All introduce
further hypotheses about the early universe or the structure of the
electroweak interaction \cite{b}. All aim to generate fields by causal
processes when the Universe is of finite age. Therefore, any magnetic field
created by these means will exist only on very small scales with an energy
density that is a negligible fraction of the background equilibrium
radiation energy density.

Nevertheless, while such fields might still provide the seeds for non-linear
dynamos in the post-recombination era, any {\it large-scale} primordial
magnetic field with a strength of order $B\simeq 10^{-8}$ Gauss, comparable
to that inferred from the lowest measured intergalactic fields and close to
the observational upper limits via Faraday rotation measurements \cite{x},
may well be of cosmological origin. A similar pregalactic (or protogalactic)
field strength is inferred from the detection of fields of order $10^{-6}$
Gauss in high redshift galaxies \cite{y} and in damped Lyman alpha clouds 
\cite{z}, where the observed fields are likely to have been adiabatically
amplified during protogalactic collapse. In the absence of a plausible
dynamo for generating large-scale pregalactic fields, it is of interest to
reconsider the limits on a large-scale primordial field in view of new
observational constraints that we outline below.

Primordial magnetic fields can leave observable traces of their influence on
the expansion dynamics of the Universe because they create anisotropic
pressures and these pressures require an anisotropic gravitational field to
support them. Primordial nucleosynthesis constraints only limit the
equivalent current epoch field to be less than about $3\times 10^{-7}$ Gauss 
\cite{e}, a value that is only slightly stronger than the dynamical
constraint at nucleosynthesis \cite{d}. We show in this letter that the
cosmic microwave background isotropy provides a stronger limit on the
strength of a homogeneous component of a primordial magnetic field.

We consider the cosmological evolution of the most general homogeneous
magnetic fields, calculate their gravitational effects on the temperature
anisotropy of the microwave background radiation, and hence derive a strong
limit on the strength of any homogeneous cosmological magnetic field by
using the 4-year Cosmic Background Explorer (COBE) microwave background
isotropy measurements \cite{l}. We employ statistical sampling techniques
appropriate for the non-Gaussian statistics of the large-scale temperature
anisotropy pattern created by a general homogeneous cosmological magnetic
field and allow for the randomness of the angle at which COBE views the
characteristic anisotropy pattern on the sky \cite{f}. The addition of a
homogeneous cosmological electric field will not be considered: a
homogeneous intergalactic electric field would create a current of charged
particles, and rapidly decay.


Our limit derives from the non-linear coupled evolution of the shear
anisotropy and magnetic field density during the radiation era which we
shall discuss below. In the presence of an equilibrium background of
blackbody radiation with isotropic momentum distribution, pressure $p_r,$
density $\rho _r,$ and equation of state $p_r={\frac 13}\rho _r$, the
anisotropic magnetic pressure prevents the rapid decay of the expansion
shear anisotropy familiar in Bianchi type I universes containing only matter
with isotropic pressure. An approximate analysis of the problem in an
axisymmetric magnetic Bianchi type I universe was given by Zeldovich \cite{g}%
. The isotropic expansion is stable at second-order and the anisotropies
decay only logarithmically in time relative to the mean expansion rate.
Physically, the evolution of the anisotropy is governed by the magnetic
pressure anisotropy.

We need to describe the evolution of the magnetic field strength and the
accompanying shear anisotropy, which will distort the microwave background
isotropy, in the most general anisotropic universes possessing shear and
3-curvature anisotropies. From earlier studies of the evolution of the most
general homogeneous universes in the absence of magnetic fields \cite{h,i},
the pattern of evolution of magnetic universes during the radiation era can
be deduced. In general, if the universe contains a non-interacting mixture
of isotropic blackbody radiation and any matter source possessing an
energy-momentum tensor with zero trace (e.g. magnetic or electric fields,
long-wavelength gravitational waves, or an anisotropic distribution of
collisionless massless or relativistic particles \cite{n}), then the
evolution follows the same characteristic pattern. The most general
anisotropic flat and open universes containing isotropic universes are of
Bianchi type VII and are equivalent to simpler flat or open anisotropic
spacetimes of Bianchi types I or V to which gravitational waves have been
added. The Einstein tensor for the more general models can be split into two
pieces: one corresponding to a simpler Bianchi type universe, the other to
an effective energy-momentum tensor for the gravitational waves which has
zero trace \cite{h,i}. This means that the anisotropies in simple Bianchi
type I (flat) or type V (open) universes containing blackbody radiation plus
trace-free matter sources with anisotropic pressures have similar
time-evolution to the most general anisotropic universes of type VII
containing blackbody radiation and a magnetic field: the 3-curvature
anisotropies and the magnetic stresses display the same asymptotic time
evolution. When the deviations from isotropy are small, the evolution of the
shear and the energy density of any trace-free matter field with anisotropic
pressures is well-approximated by setting the blackbody density and the
volume-averaged Hubble expansion rate equal to their values in the isotropic
flat Friedmann universe ($\rho _r={\frac 3{4t^2}}$ and ${H={\frac 1{2t}},}$
since we set $8\pi G=1$) before solving for the non-linear evolution of the
shear anisotropy and the magnetic field. The ratio of the shear to Hubble
expansion rate has a generic behaviour for $p\leq {\frac 13}\rho $; (when $p>%
{\frac 13}\rho $ the anisotropic stresses dominate at large $t$ and the
solution ceases to be a small perturbation of an isotropic Friedmann
universe as $t\rightarrow \infty $). In general, when anisotropies are
small, the ratio of the shear anisotropy, $\sigma $, to the mean Hubble
rate, $H$, relaxes to a constant value determined by the ratio of the total
energy densities in anisotropic traceless fluids and 3-curvature
anisotropies, $\rho _{ga},$ and magnetic fields, $\rho _B,$ to that of the
isotropic perfect fluid density, $\rho ,$ (radiation or dust). When the
isotropic fluid has equation of state $p=(\gamma -1)\rho $, with $0$ $%
<\gamma \leq 4/3,$ the time evolution is determined by the Einstein equations

\begin{eqnarray}
\frac d{dt}(\frac \sigma H)^{} &=&(\frac \sigma H)\left( \frac{\gamma -2}{%
\gamma t}\right) +\frac 4{\gamma t}\left( \frac{\rho _{ga}+\rho _B}\rho
\right) ,  \label{evoln1} \\
\frac d{dt}\left( \frac{\rho _{ga}+\rho _B}\rho \right) &=&-\frac 2{9\gamma t%
}\left( \frac{\rho _{ga}+\rho _B}\rho \right) \left( 4\frac \sigma H+9\gamma
-12\right) .  \nonumber \\
&&  \nonumber
\end{eqnarray}
Here, $\rho _B=B^2/8\pi $ is the magnetic field density. Hence, as $%
t\rightarrow \infty ,$ we have $\sigma /H\rightarrow $ constant and

\begin{eqnarray}
\frac d{dt}\left( \frac{\rho _{ga}+\rho _B}\rho \right)  &=&\frac 2{9\gamma t%
}\left( \frac{\rho _{ga}+\rho _B}\rho \right)   \nonumber \\
&&\times \left[ \left( \frac{16}{\gamma -2}\right) \left( \frac{\rho
_{ga}+\rho _B}\rho \right) +9\gamma -12\right]   \label{evoln2}
\end{eqnarray}
In the radiation era ($\rho =\rho _r,\gamma =4/3),$ at redshifts $%
1+z>1+z_{eq}=8\times 10^3\Omega _0h_{50}^2$, where $\Omega _0\leq 1$ is the
cosmological density parameter and $h_{50}$ is the present value of the
Hubble constant in units of $50Kms^{-1}Mpc^{-1}$, we have

\begin{equation}
\rho _B/\rho _r\rightarrow Q/[1+4Q\ln (t/t_0)];\text{ }Q\text{ }constant.
\label{asympt}
\end{equation}
During the dust era ($\rho =\rho _d,\gamma =1),$ when $z<z_{eq}$, the
evolution is determined at linear order and $(\rho _{ga}+\rho _B)/\rho
_d\propto 1+z$ falls linearly with redshift. In general, the shear
distortion created by magnetic fields and any other trace-free anisotropic
stresses is given by 
\begin{eqnarray}
{\frac \sigma H}={\frac 4{{2-\gamma }}}\left( \frac{\rho _B}\rho +\frac{\rho
_{ga}}\rho \right) +\delta t^{\frac{{\gamma -2}}\gamma }  \label{maineq}
\end{eqnarray}
where $\delta $ is a constant. The $\delta $ term gives the simple shear
decay for universes with isotropic 3-curvature containing only matter with
isotropic pressure; this term becomes negligible at late times. In both dust
and radiation eras the anisotropic and magnetic stress terms dominate the $%
\delta $ term at late times and produce a slower decay of the shear
distortion. Note that the presence of curvature anisotropy or any
anisotropic trace-free matter stress changes the shear evolution from the
simple delta term that is usually studied in the literature (eg in \cite{p}%
). This behaviour can be seen in the dust era of a magnetic universe in the
exact magnetic dust solution of Thorne \cite{d} and also in the study of
almost isotropic dust plus collisionless trace-free radiation of \cite{j}.
It was employed in the study of nucleosynthesis in ref. \cite{o} in
conjunction with the early microwave background isotropy data. Also, note
that the ratio of the magnetic and blackbody radiation densities is not
constant (as assumed for example in \cite{e} and \cite{q}), but falls
logarithmically during the radiation era, and the three expansion
scale-factors, $a(t),b(t),c(t)$ evolve with time, to first order in the
anisotropy, with $a(t)\propto b(t)\propto t^{1/2}\{\ln (t/t_0)\}^{\frac 14}$
and $c(t)\propto t^{1/2}\{\ln (t/t_0)\}^{-\frac 12},$ with $H=(2t)^{-1}$ as
assumed at leading order. During the dust era we have $a(t)\propto
b(t)\propto t^{2/3}\{1-Kt^{-2/3}\}$ and $c(t)\propto t^{2/3}\{1+2Kt^{-2/3}\},
$ with $K>0$ a constant such that $\rho _B/\rho _d=\frac 43Kt^{-2/3}$ and $%
H=\ 2/(3t)$ first order.

The magnetic field and the accompanying shear distort the microwave
background temperature isotropy in accord with the most general anisotropic
universes of Bianchi type VII. By combining eqn.(\ref{maineq}) and the
expression for the angular anisotropy pattern in type VII, \cite{k}. 
\begin{eqnarray}
{\Delta T}_A({\bf {\hat r}})=({\frac \sigma H})_{ls}Y(\theta ,\phi ,\Omega
_0,h_{50},x,z_{ls})  \label{bia}
\end{eqnarray}
where {\it `ls'} denotes the epoch of last scattering ($1+z_{ls}=1100$ with
no reheating but if there is reionization of the universe then last
scattering can be delayed until $1+z_{ls}=78(\Omega _bh_{50})^{-1}$ for $%
\Omega _0z_{ls}<<1,$\cite{u}) where $\Omega _b$ is the baryon density
parameter. The exact form of the pattern function $Y$ can be read off from
Eqs. (4.11)-(4.16) of Barrow, Juszkiewicz and Sonoda\cite{k}. The constant
parameter $x$, introduced by Collins and Hawking \cite{h} is a measure of
the three-curvature anisotropy configuration in the type VII universes. It
contributes a spiral with $2/\pi x$ twists to the angular pattern, \cite{k}.
If $\Omega _0<1$ there will be a focusing the quadrupole towards the axis of
anisotropy, generating a hot spot.

The problem of constraining global anisotropy is substantially different
from the traditional statistical task of estimating parameters in Gaussian
models. In the latter case, the ensemble is entirely characterized by the
power spectrum while in the former, a given set of parameters corresponds to
a completely specified pattern in the sky, up to an arbitrary rotation. This
problem was dealt with in some detail in \cite{f}. One can model the
microwave background signal as the sum of two components: a {\it %
statistically isotropic} Gaussian random field $\Delta T_I$, which we
assumed to have a scale invariant power spectrum on the scales we are
interested in, and a {\it global}, {\it anisotropic} pattern, $\Delta T_A$,
as in eq. (\ref{bia}), which is uniquely defined by the set of parameters $x$%
, $\Omega _0$, $(\sigma /H)_0,$ and $\theta $, $\phi $ (its orientation on
the sky). Each pixel of the data set is given by $d_i=(\Delta T\star \beta )(%
{\bf r}_i)+N_i$ where $\beta $ represents the DMR beam pattern, ${\bf r}_i$
is a unit vector pointing in the direction of pixel $i$ and $N_i$ is the
noise in pixel $i$. To an excellent approximation, one can assume that $N$
is Gaussian ``white'' noise, i.e. $\langle N_iN_j\rangle =s_i^2\delta _{ij}$.

Our task is, given a pair ($x$, $\Omega _0$), to find the orientation ($%
\theta $, $\phi $) which allows the maximum observed value of $(\sigma /H)_0$%
. One can do this using standard frequentist statistical methods: we define
a goodness-of-fit statistic that depends on the data, compute its value for
the actual data, and then compute the probability that a random data set
would have given a value as good as the actual data. In \cite{f} $\eta $ was
defined to be: 
\begin{eqnarray}
&&  \nonumber \\
\eta &=&\min_{\sigma ,\theta ,\phi }\eta _1;\text{ where }\eta _1={\frac{%
\Delta _0^2-\Delta _1^2}{\Delta _0^2};}
\end{eqnarray}
$\Delta _0^2$ is the noise-weighted mean-square value of the data and $%
\Delta _1^2$ is the noise-weighted mean-square value of the residuals after
we have subtracted off the anisotropic part. Note that removing the
incorrect anisotropic portion will only increase the residuals so the
difference between the two terms is an obvious choice for a goodness-of-fit.
Dividing by $\Delta _0^2$ ensures a weak dependence on the amplitude of the
isotropic component, while defining the statistic as the minimum of $\eta _1$
allows us to deal with the uncertainty in ($\theta ,\phi $).

This statistical method was applied to the 4-yr COBE DMR data-set. The two
53 GHz and the two 90 GHz maps were averaged together, each pixel weighted
by the inverse square of the noise level, to reduce the noise level in the
average map. All pixels within the Galactic cut were removed so as to reduce
Galactic contamination, and a best-fit monopole and dipole were subtracted
out. The map was degraded from pixelization 6 to pixelization 5 (i.e.
binning pixels in groups of four). Simulations were performed for a set of
models from the ($\Omega _0,x$) plane; for each choice of the three
parameters ($\Omega _0$, $x$ and $(\sigma /H)_0$) approximately 200 to 500
random DMR sets were generated, so allowing us to determine an approximate
fit to the probability distribution function of $\eta $.

The most conservative limit on the cosmological magnetic field arises when
we assume $\rho _{ga}<\rho _B$ so the whole anisotropy is contributed by the
magnetic field stresses. A reasonable fit of the upper bound at a $95\%$
confidence level is: 
\begin{eqnarray}
({\frac{\rho _B}\rho })_0 &<&{\frac{{(2-\gamma )}}4}f(x,\Omega _0)\times
10^{-9}  \nonumber \\
f &\simeq& 2.1 \ \Omega_0^{.33}x^{-.01/\Omega_0}  \nonumber \\
& &\mbox{ with $x\in[.01,3]$ and $\Omega_0\in[.1,1],$}
\end{eqnarray}
where we have considered the largest possible contribution from the magnetic
component. Note that the ``shape'' factor is roughly bounded by $0.6$ $<f<$ $%
2.2$. This gives us a final bound on the magnitude of the magnetic field
today of 
\begin{eqnarray}
B_0<3.5\times 10^{-9}f^{\frac 12}(\Omega_0h_{50}^2)^{\frac 12}{\rm Gauss}.
\end{eqnarray}
This bound can be improved by a factor of $\sqrt{3}$ if one considers the
results from \cite{s}. In this case, a slightly different goodness-of-fit
statistic is used: instead of working with the noise-weighted quantities, $%
\Delta _0^2$ and $\Delta _1^2$, the authors chose to weight the pixels with
the covariance matrix of the total Gaussian components (i.e. the noise and
isotropic cosmological components). This gives a limit of

\begin{equation}
B_0<2.3\times 10^{-9}f^{\frac 12}\ (\Omega _0h_{50}^2)^{\frac 12}{\rm Gauss}
\end{equation}
Note that the microwave background limits on the amplitude of anisotropies
are much stronger than those imposed by nucleosynthesis \cite{o}. In
unrealistic models with no anisotropic matter stresses (and therefore no
magnetic field) and isotropic curvature, the shear falls off rapidly in
accord with the $\delta $ term in (\ref{maineq}) and the limits from
nucleosynthesis would be stronger. But with anisotropic matter stresses,
magnetic fields, or anisotropic curvature, the anisotropy falls only
logarithmically during the radiation era. The limits on $\rho _B/\rho _r$ at
nucleosynthesis 
are only $O(1)$ and the log decay means they are weaker than limits $%
O(10^{-5})$ imposed at $z=1.1\times 10^3$ by the microwave background. If
there is reheating and last scattering occurs at $z<<1100$ then the analysis
is slightly changed but last scattering would need to occur at a redshift
lower than $6\Omega _0h_{50}^2$ for the nucleosynthesis limit to be
competitive with the microwave limit. This never happens.

Adams et al \cite{q} have used the weak nucleosynthesis limit of Grasso and
Rubinstein of $B_0\leq 3\times 10^{-7}$ Gauss obtained for random fields to
argue that a cosmological magnetic field could lead to observable
distortions of the acoustic peaks in the microwave background. Our limit on $%
B_0$ rules out any observable effect on the acoustic peaks. In fact, when
the $(\ln t)^{-1}$ decay of $\rho _B/\rho _r$ of eqn.(\ref{asympt}) is taken
into account a nucleosynthesis limit is strong enough to render the acoustic
peak distortions unobservable. Our limit permits a field strength of $%
10^{-9} $ Gauss required to induce a measurable Faraday rotation in the
polarization of the microwave background \cite{r}.

Any period of inflation long enough to explain the horizon and flatness
problems would necessarily reduce magnetic field effects and their
associated anisotropies to unobservably low levels. If $N$ e-folds of de
Sitter inflation occurs ($p=-\rho =$constant) then $\sigma /H$ will be
reduced by $\exp (-3N)$ and $\rho _B/\rho $ will be reduced by a factor $%
\exp (-4N)$ and $N\sim 70$ is sufficient to solve the horizon problem \cite
{t}. The formula (\ref{maineq}) for $\sigma /H$ applies to the case of
inflation if the $\delta $ term is changed to $\delta \exp (-3N)$. Note that
if generalized inflation occurs ($0\leq \gamma <2/3$), the $\delta $ term of
eqn. (\ref{maineq}) always dominates the $(\rho _{ga}+\rho _B)/\rho $ term
as $t\rightarrow \infty $ unlike in the non-inflationary case when $%
2/3<\gamma <4/3$. All anisotropies decay in accord with the no-hair theorem
when the curvature is non-positive because the anisotropic trace-free
stresses obey the strong energy condition\cite{m}.  The discovery of
microwave background patterns characteristic of large-scale homogeneous
anisotropy or a homogeneous primordial magnetic fields in future
observational programs would therefore rule out the standard picture of
inflation.

{\it Acknowledgements}: We acknowledge partial support by NASA and DOE. P.F.
was supported by the Center for Particle Astrophysics, a NSF Science and
Technology Center at U.C.~Berkeley, under Cooperative Agreement No. AST
9120005. J.D.B. was also supported by the PPARC.

\pagebreak
\pagestyle{empty}

\end{document}